# Generation of ultrashort light pulses carrying orbital angular momentum using a vortex plate retarder-based approach

Tlek Tapani[1,¶], Haifeng Lin[1,¶], Aitor De Andres[1], Spencer W. Jolly[2], Hinduja Bhuvanendran[1], and Nicolò Maccaferri[1*]

[1]Department of Physics, Umeå University, Umeå, Sweden
[2]Service OPERA-Photonique, Université libre de Bruxelles, Bruxelles, Belgium
*nicolo.maccaferri@umu.se
[¶]Contributed equally

**Abstract**
We use a vortex retarder-based approach to generate few optical cycles light pulses carrying orbital angular momentum (known also as twisted light or optical vortex) from a Yb:KGW oscillator pumping a noncollinear optical parametric amplifier generating sub-10 fs linearly polarized light pulses in the near infrared spectral range (central wavelength 850 nm). We characterize such vortices both spatially and temporally by using astigmatic imaging technique and second harmonic generation-based frequency resolved optical gating, respectively. The generation of optical vortices is analyzed, and its structure reconstructed by estimating the spatio-spectral field and Fourier transforming it into the temporal domain. As a proof of concept, we show that we can also generate sub-20 fs light pulses carrying orbital angular momentum and with arbitrary polarization on the first-order Poincaré sphere.

**1. Introduction**
Light carrying orbital angular momentum (OAM), known also as twisted light or optical vortex, is characterized by a field profile with a phase singularity, such that the phase is undefined in the center of the vortex and thus the intensity is zero, and the phase variation along a path enclosing the singularity is non-zero[1–4]. Noteworthy, optical vortex beams have led to many practical applications[5], such as the detection of astronomical objects or incoherent light with signal enhancements of several orders of magnitude[6], the breaking of the diffraction resolution limit in fluorescence microscopy[7], as well as to probe magnetization states in matter[8] or image a variety of intrinsic and extrinsic properties encoded in phase and amplitude gradients and dislocations of a complex electromagnetic field[9]. The most commonly used methods to convert a Gaussian laser beam into a vortex beam are based on phase masks[10], refractive-diffractive elements[11,12] or diffraction holograms[13]. Other methods include liquid crystal-based order electrically controlled q-plate systems[14], nanoscale collective electronic excitations known as plasmons[15,16], and spiral phase plates[17], although it is difficult to assess the impact of the inherent chromaticity of the phase plate has on the quality of (broadband) ultrashort vortices. On one hand, when focusing on continuous wave light beams or not interested in either the spatio-temporal structure and coupling or the temporal duration of pulsed light, these methods are extremely beneficial. On the other hand, generating ultrashort vortices is challenging since most techniques used to create monochromatic vortices have limitations when applied to broadband light sources[18–20]. Even though polychromatic methods have been suggested and applied to ultrashort pulse vortex generation[21–24], these approaches are quite complex to implement, making monochromatic techniques still attractive but limited to the application to continuous wave beams or long pulses, where the variation of the pulse temporal duration is not an important factor. In this work, we create sub-20 fs optical vortices by using a vortex half-wave retarder (VR) plate approach. The VR plate can generate OAM beams similarly to electrically activated liquid crystal q-plates, but it does not require electrical input to be driven as q-plates do. We characterize our light pulses carrying OAM both spatially and temporally by using the astigmatic imaging technique and the second harmonic generation-based frequency resolved optical gating (SH-FROG), respectively. We then analyze the generation of the optical vortices and reconstruct their spatio-temporal structure. As a proof of concept, we also provide experimental evidence, which is corroborated by simulations, that with our approach we can also generate light pulses carrying both OAM and an arbitrary polarization state on the first-order Poincaré sphere.

## 2. Experimental setup

Our setup is based on a commercial Yb:KGW regenerative amplifier (PHAROS, Light Conversion), operating at a repetition rate of 50 kHz. This laser pumps a commercial optical parametric amplification (OPA) system, a custom ORPHEUS-N-2H, which exploits the second harmonic of the fundamental wavelength to generate broadband light pulses in the range 700-900 nm. By using a commercial pulse compressor, the OPA system outputs vertically polarized light pulses with a sub-10 fs time duration, centered at a wavelength of 850 nm and carrying an energy of 10 µJ per pulse. These pulses undergo the process of acquiring OAM, as we insert optical components, such as half-waveplates (HWPs), quarter-waveplates (QWPs), and vortex plates along the beamline (see Figure 1), ensuring precise control over the spin and orbital optical states (helicity $s$ and OAM $l$, respectively) of the pulses. Two QWPs enable to control the spin angular momentum carried by the light pulse ($s$), where $s = \pm 1$ refers to the photon helicity that is related to left/right-handed circularly and $s = 0$ refers to a linearly polarized state. We can add OAM to the light pulses by inserting a vortex half-wave retarder (VR) plate (VRA-850, LBTEK) between the waveplates, where the integer number $l$ refers to the topological charge of the optical vortex. The VR plate is made of N-BK7 glass substrates and liquid crystal polymers (LCP) material. It is a sandwich structure of the LCP layer and the front and rear glass substrates. It has a constant half-wave retardance across the clear aperture, but the fast axis orientation of the liquid crystal molecules rotates continuously over the area of the optical element. First, we insert an achromatic QWP, followed by VR plate, and ending with a second achromatic QWP. The initial QWP is utilized to convert the vertically polarized pump beam into a circularly polarized light beam. Subsequently, the VR plate is employed to generate a vortex beam with circular polarization. Finally, the last QWP is utilized to convert the circularly polarized vortex beam into a linearly polarized vortex beam. In this way we can generate linearly polarized light carrying only OAM.

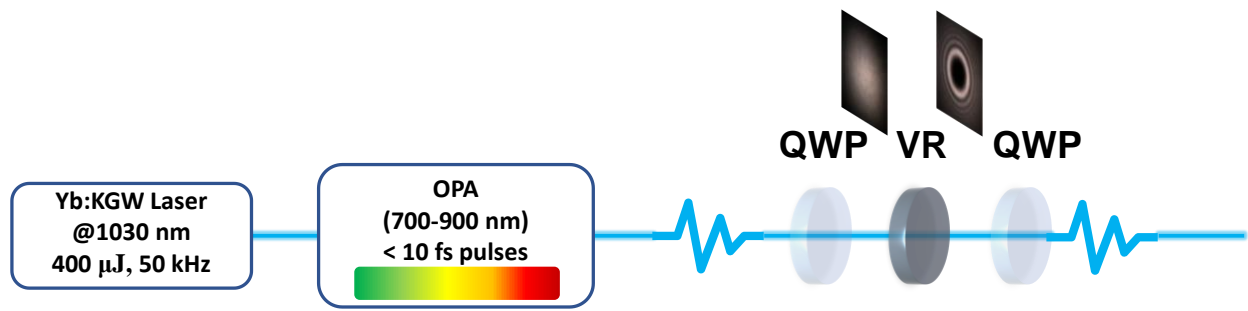

**Figure 1.** Sketch of the setup used to generate twisted light pulses.

The temporal characterization of the OPA output pulses before adding to the OAM is carried out by using the second-harmonic generation frequency-resolved optical gating (SHG-FROG) technique[25]. Figure 2a and 2b show the measured and reconstructed FROG trace of the initial OPA output pulse with a total retrieval error of <0.01%. Figure 2c shows the experimental pulse spectrum (blue curve) centered at 850 nm with the reconstructed spectral phase. The retrieved time profile (dotted blue curve), along with its corresponding phase (red curve), is displayed in Figure 2d. The Gaussian fit (solid blue curve in Figure 2d) provides a pulse duration of 9.6 fs (calculated as the FWHM of the Gaussian fit). After this preliminary characterization, the light pulses are ready to acquire OAM and being further characterized.

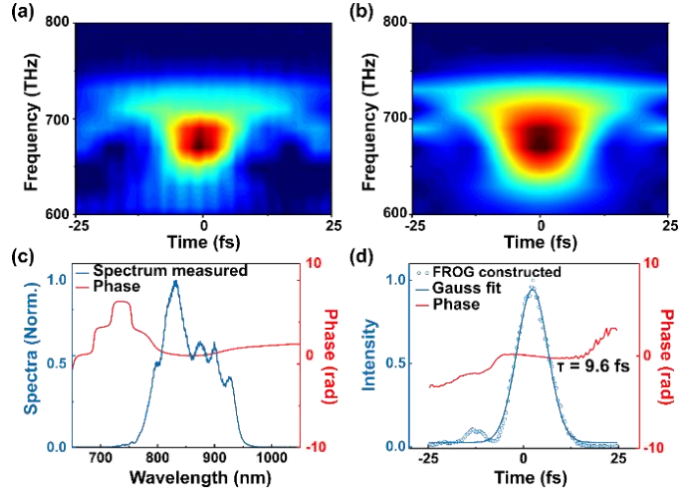

**Figure 2.** FROG characterization of the compressed 2H pulses. (a) measured and (b) reconstructed OPA output pulse FROG traces; (c) OPA output pulse spectrum with the reconstructed spectral phase; (d) reconstructed OPA output pulse profile with the Gaussian fit (blue solid line) and corresponding temporal phase.

### 3. Generation of light pulses carrying OAM

As shown in Figure 3a, the polarization state of the light pulse at each position (1-4) is analyzed by using an ultra-broadband wire grid polarizer (WGP) covering the wavelength range 250 - 4000 nm, in combination with a power meter. The initial state of polarization (SOP) is linear along the vertical direction (grey dots), which we verified with a polarizer-analyzer, obtaining an extinction ratio of 16 dB. We observe that after passing through the achromatic QWP, which is supposed to have its fast axis with respect to vertical direction at 45°, the SOP becomes slightly elliptical (blue dots).

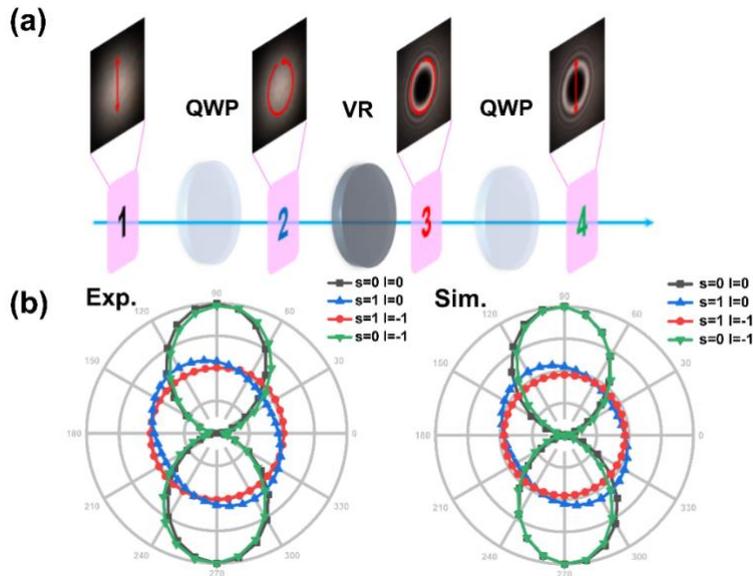

**Figure 3.** (a) Polarization states at each position 1-4. (b) Intensity polar plots of experimental results and theoretical computations as a function of polarizer angle at each position 1-4.

We ascribe this to a non-perfect linear polarization input and a slight misalignment of the fast axis of the QWP. To verify this assumption, using the Jones matrix formalism we simulated the beam intensities evolution after each optical component, with the parameters of input SOP of [-0.05+0.03i 1]$^T$ and QWP at 44.5°. Comparing with the ideal case of SOP of [0 1]$^T$ and QWP at 45°, as shown on the right panel in Figure 3b, the experimental results are well reproduced by our calculations with the artificial deviation. It is worth noting that the SOP turn into more circular after the VR plate (red dots), which has the spatial distribution of fast axis of liquid crystal, as confirmed also by the simulations. Finally, after passing through a second QWP, the SOP comes back to linearly polarized, but this time carrying OAM (green dots).

## 4. Spatial and temporal characterization of light pulses carrying OAM

Figure 4 shows the intensity profiles of the generated linearly polarized light pulses with different topological charges ($s = 0, l = 0,1,2,3,4$), measured with a complementary metal-oxide-semiconductor (CMOS) camera. A region of nearly null intensity (corresponding to the phase singularity in the optical vortex) is observed in the center of the beam profile, with a size that increases with the increasing value of OAM number $l$. The spatial characterization of the vortex pulses has been carried out by focusing the beam directly on a CMOS camera with a tilted concave mirror, which introduced astigmatism. The number of dark lines between the end nodes indicates the OAM value, that is $l$, of the vortex wavefront. The imaging is performed with full pulse bandwidth and at wavelengths of 800 nm, 850 nm, and 900 nm. The consistent outcomes observed across the full pulse spectrum and its individual colors affirm the presence of the same OAM across the entire bandwidth.

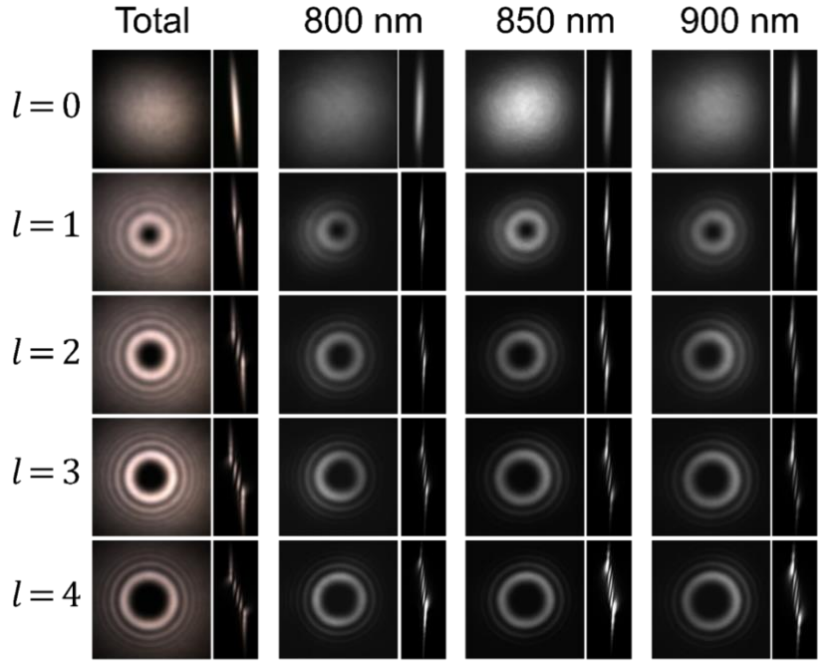

**Figure 4.** Intensity profiles of the near infrared pulses ($s = 0, l = 0,1,2,4$) with entire pulse bandwidth (total) and at wavelengths of 800 nm, 850 nm, and 900 nm. The topological charge of these beams can be differentiated by imaging the intensity at the focal region of a tilted concave mirror or an astigmatic lens, as shown in the right side of each intensity profile.

The pulses were temporally characterized for each value of OAM order ($l = 0,1,2,3,4$) by sending the beam to a SHG-FROG setup. The measurement for the case where $l = 0$ is conducted using the same configuration as those where $l = 1,2,3,4$, with the only modification being the replacement of the VR plate with a glass plate of comparable thickness (3 mm). Despite the diversity in OAM order (l=0,1,2,3,4) across the tested pulses, all of them exhibited a consistent temporal duration of 12-13 fs. Figure 5 below plots the FROG traces and analysis of them for pulses with $l = 0,1,4$. The pulse duration has increased beyond that of the initial 2H OPA output light pulse due to its passage through additional optical components. Even though the additional group delay dispersion was pre-compensated by readjusting our chirped-mirror compressor in each case, the higher-order dispersion components could not be fully addressed. Noteworthy, if used in pump-probe experiments, the uniformity in the time duration among these pulses enables a direct comparison of experimental results obtained by pumping a system with pulses carrying different OAM orders.

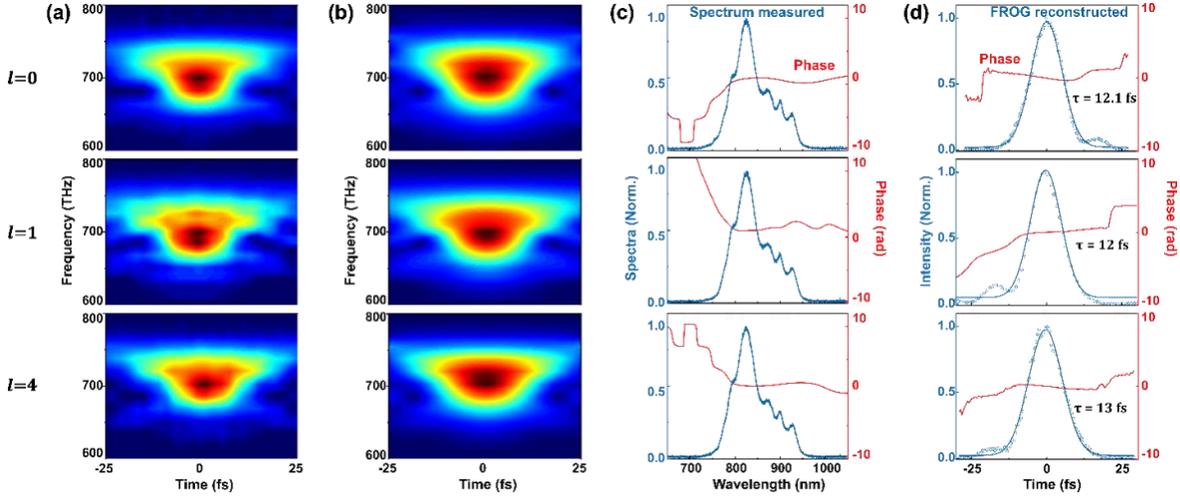

**Figure 5.** FROG characterization of the light pulses carrying distinct OAM ($l = 0,1,4$). (a) measured and (b) reconstructed FROG traces; (c) pulse spectra with the reconstructed spectral phase; (d) reconstructed pulse profile with the blue solid line showing the Gauss fit curve and the corresponding temporal phase.

## 5. Generation of light pulses carrying OAM and arbitrary polarization

The orientation of optical axis of the VR plate can be expressed as:

$$\alpha = \frac{m\varphi}{2} + \varphi_0 \tag{1}$$

where $m$ is an integer number, $\varphi = \arctan(y/x)$ is the azimuthal angle, and $\varphi_0$ is the initial fast axis orientation on the wavefront plane. $\varphi_0$ equals to 0 when we set it along the $x$ axis. The Jones matrix of the VR plate can be written as:

$$M = \begin{bmatrix} \cos m\varphi & \sin m\varphi \\ \sin m\varphi & -\cos m\varphi \end{bmatrix} \tag{2}$$

It is known that a general homogeneous polarization $P_0$ can be expressed in terms of its polar angle $\varepsilon'$ and azimuthal angle $\eta'$ on the Poincaré sphere (PS)[26–28]:

$$\vec{P_0} = \sin\left(\varepsilon' + \frac{\pi}{4}\right)|\vec{R}\rangle + \exp(2i\eta')\cos\left(\varepsilon' + \frac{\pi}{4}\right)|\vec{L}\rangle \tag{3}$$

where $|\vec{R}\rangle = (\vec{x} + i\vec{y})/\sqrt{2}$ and $|\vec{L}\rangle = (\vec{x} - i\vec{y})/\sqrt{2}$ represent the right and left circular polarization components, respectively. Similarly, the SOP of any point on the high-order PS can be represented by a linear superposition of two poles $|\overrightarrow{R_m}\rangle$ and $|\overrightarrow{L_m}\rangle$, where $|\overrightarrow{R_m}\rangle = \exp(-im\varphi)(\vec{x} + i\vec{y})/\sqrt{2}$ and $|\overrightarrow{L_m}\rangle = \exp(im\varphi)(\vec{x} - i\vec{y})/\sqrt{2}$. The two poles on the high-order PS denote the orthogonal circular polarization basis with opposite topological charges. Consequently, the output beams from the VR plate can be calculated by:

$$\begin{aligned}M\overrightarrow{P_0} &= \exp(-im\varphi)\cos\left(\varepsilon' + \frac{\pi}{4}\right)|\vec{R}\rangle + \exp(im\varphi - 2i\eta')\sin\left(\varepsilon' + \frac{\pi}{4}\right)|\vec{L}\rangle \\ &= \sin\left(\frac{\pi}{4} - \varepsilon'\right)|\overrightarrow{R_m}\rangle + \exp(2i(\pi - \eta'))\cos\left(\varepsilon' + \frac{\pi}{4}\right)|\overrightarrow{L_m}\rangle\end{aligned} \quad (4)$$

In contrast to Eq. (3), a direct mapping is established between the conventional PS and the high-order PS via the VR plate, and their angular coordinates exhibit the following correlation:

$$\eta = \pi - \eta' \quad (5)$$

$$\varepsilon = -\varepsilon' \quad (6)$$

As example, here we focus on first-order PS by employing the VR plate with the order $m = 1$. The experiments are conducted to validate the aforementioned principles[26–28], as illustrated in Figure 6a. We generate the vector beams on the points A, B, C and D on the first-order PS by passing through the linear phase retarder (achromatic HWP or QWP) and VR, as shown in Figure 6b. First, for the vector beams on the point A and B, we set the fast axis of an achromatic HWP with respect with $x$ axis (θ) at 0° and 22.5°, respectively. Next, we replace the achromatic HWP by an achromatic QWP and set its fast axis (θ') at -45° and 22.5°, respectively, to achieve the vector beams on the point C and D. The initial SOP is $[1\ 0]^T$, which is linearly polarized along the $x$ axis. For the A and B cases located on the equator of the PS, their polar angle $\varepsilon = 0$ and azimuthal angle $\eta = \pi - 2\theta$. Moreover, the Stokes parameters after the achromatic QWP (points C and D) on conventional PS were calculated to be $[1\ \cos^2(2\theta')\ \cos(2\theta')\sin(2\theta')\ \sin(2\theta')]^T$, resulting in the polar angle $\varepsilon = -\theta'$ and azimuthal angle $\eta = \pi - \theta'$ on the first-order PS due to the Eqs. (5) and (6).

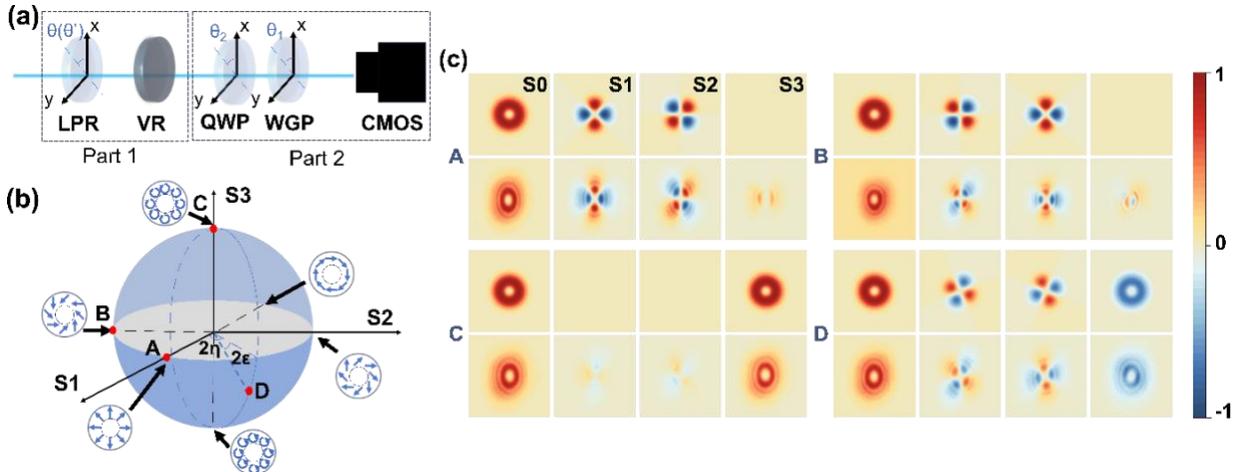

**Figure 6.** (a) The experimental setup for generating first-order PS beams (Part 1) and measuring first-order PSs (Part 2). LPR, linear phase retarder, either HWP or QWP; (b) Schematic illustration of the first order PS. Insets A, B, C and D points are the local polarization vector states at various positions on the sphere. (c) Stokes parameters (S0-3) of the generated vector beams corresponding to A, B, C and D points on the first order PS. For each point, the first row shows results from theory, and second row shows the results from the experiment.

A typical setup (QWP, WGP, and CMOS camera) measures the Stokes parameters that describe the SOP and intensity of the output light. The Stokes parameters $S_0$, $S_1$, $S_2$, and $S_3$ are given by:

$$\begin{aligned} S_0 &= I(0°, 0°) + I(90°, 90°) \\ S_1 &= I(0°, 0°) - I(90°, 90°) \\ S_2 &= I(45°, 45°) - I(135°, 135°) \\ S_3 &= I(-45°, 0°) - I(45°, 0°) \end{aligned} \tag{7}$$

Here $I(\theta_1, \theta_2)$ is the intensity recorded by the CMOS camera when the polarization direction of the WGP and the optical axis of QWP stay at $\theta_1$ and $\theta_2$ against the $x$ axis, respectively. As shown in Figure 6c, the experimental results (the bottom rows) agree well with the theoretical calculations (the top rows) at each point. In principle, we can obtain arbitrary cylindrical vector beams on the first-order PS by employing a combination of QWP and HWP. By slightly readjusting the pulse compressor, we are able to attain a time duration of 11.9 femtoseconds for the generated D points (see Figure 7), similar to the previously characterized vortex pulse results. While not explicitly illustrated, a time duration around 12 femtoseconds has also been measured for points A, B, and C. This proves that we can generate ultrashort light pulses carrying OAM at all points on the spherical spectrum represented in Figure 6b.

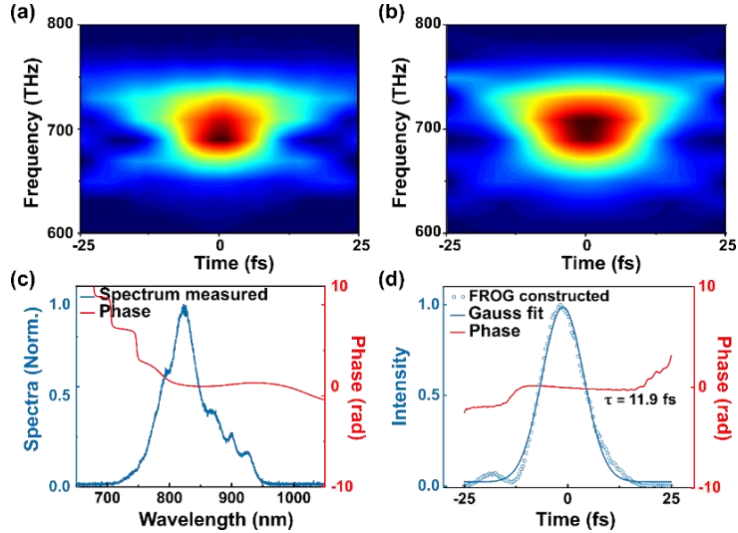

**Figure 7.** FROG characterization of the D point on the first order PS (Fig 6a). (a) measured and (b) reconstructed FROG traces; (c) pulse spectra with the reconstructed spectral phase; (d) reconstructed pulse profile with the blue solid line showing the Gauss fit curve and the corresponding temporal phase.

## 6. Reconstruction of the spatio-temporal structure of the vortices

The spatiotemporal structure of the vortices, $E(x, y, t)$, was reconstructed by estimating the spatio-spectral field, $\tilde{E}(x, y, \omega)$, and Fourier transforming it as

$$E(x, y, t) = \int_{-\infty}^{\infty} \tilde{E}(x, y, \omega) e^{-i\omega t} d\omega, \tag{8}$$

where $x$ and $y$ are the spatial coordinates transverse to the pulse propagation direction. A complete spatiotemporal characterization requires the simultaneous knowledge of both amplitude $A$ and phase $\varphi$ components in either temporal or spectral domain[19,29,20]. This can be accomplished by measuring the spatially resolved spectrum $S(x, y, \omega)$, the spectrally resolved wavefront $\theta(x, y, \omega)$ (also referred to as the spatio-spectral phase), and the spectral phase $\phi(x_0, y_0, \omega)$ at a known $x_0, y_0$ location. The latter term can be retrieved by using conventional techniques such as FROG or spectral phase interferometry for direct electric-field reconstruction (SPIDER)[30]. Then, the field is obtained by combining them as

$$\tilde{E}(x, y, \omega) = A(x, y, \omega)e^{\varphi(x,y,\omega)} = \sqrt{S(x, y, \omega)}e^{\theta(x,y,\omega) - \theta(x_0,y_0,\omega) + \phi(x_0,y_0,\omega)}, \tag{9}$$

where the spatio-spectral phase $\varphi(x, y, \omega)$ is calculated in the second step by stitching the first order phase component (the wavefront $\theta$) with the higher phase orders ($\phi$) given by the temporal measurement. Notice that this approach is typically blind to a constant phase component, the carrier envelope phase (CEP), although it could in principle be separately measured with a zeroth order phase detector and stitched in the same way, provided that the light driver is CEP stable. Nevertheless, in this work we followed a minimalistic approach and estimated the spectral field by measuring the frequency-integrated spatial-amplitude $A(x, y, \bar{\omega})$ with a beam profiler, the spatially-integrated spectrum $S(\bar{x}, \bar{y}, \omega)$ with a spectrometer, the spatially-integrated spectral phase $\phi(\bar{x}, \bar{y}, \omega)$ with SHG-FROG, and simulated a perfect wavefront $\theta(x, y, \omega)$ according to the specifications of the vortex plate manufacturer.

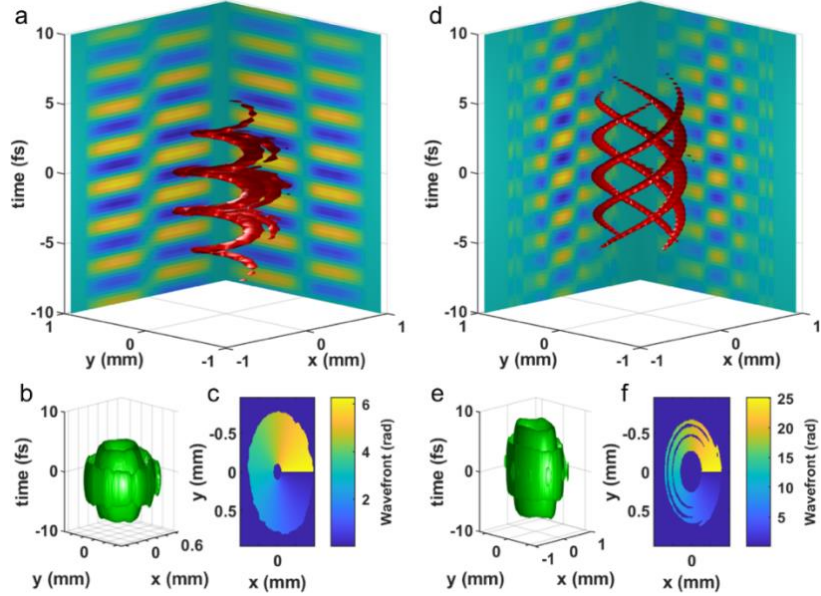

**Figure 8.** Spatiotemporal reconstruction of two vortices with OAM $l = 1$ (a-c) and $l = 4$ (d-f). (a,d) Isosurfaces of the real component of the electric field at $+\sqrt{1/2}$ of its peak value, with the integrated value along the $x$ and $y$ directions shown on the corresponding 2D maps. (b,e) Isosurfaces of the intensity at $1/2$ of its peak value. (c,f) Simulated wavefronts phase-blanked in the regions with intensity below 0.3 of the peak intensity.

Then, we reconstructed the field as explained assuming that

$$S(x, y, \omega) \approx S(\bar{x}, \bar{y}, \omega)A(x, y, \bar{\omega}), \tag{10}$$

and

$$\phi(x_0, y_0, \omega) \approx \phi(\bar{x}, \bar{y}, \omega). \tag{11}$$

It is important to notice that these approximations are only valid in the absence of spatio-temporal (or spatio-spectral) aberrations, and the result should be understood as the best-case scenario. The results of the reconstruction are displayed in Figure 8, where the spiraling structure of the pulse can be appreciated from the electric field in Figure 8a and 8d, which becomes steeper for higher angular momentum number $l$. While the local electric field frequency is approximately constant at any transversal position $(x_0, y_0)$ and across all l numbers, the magnitude of the wavefront discontinuity increases by $l \cdot 2\pi$, as can be seen in Figure 8c and 8f. Regarding the structure of the intensity, it shows the expected low intensity hole in its center along the temporal axis, as shown in Figure 8b and 8e. Spatio-temporal characterization of ultrashort vortices is a developing field[19,29,31] and may present many interesting subtleties, especially when the duration approaches few-cycle[32] or additional space-time couplings are present[33], and the platform presented here may prove useful in that context. Nevertheless, from this analysis we can see that the pulse duration is close to the original one before OAM generation.

## 7. Conclusions

We have demonstrated the possibility to generate ultrashort light pulses carrying OAM using a commercial setup based on a Yb:KGW regenerative amplifier pumping a OPA system. The spectral range of the light pulses can be shifted towards lower or larger wavelengths changing the phase matching conditions in the OPA process[34,35]. We have also shown that we can generate light pulses carrying OAM and an arbitrary polarization state on the first-order PS. We envision that this approach can be expanded as well to structure light pulses with coupled spacetime-dependent properties, including dynamic spot sizes, moving focal points, evolving orbital angular momenta and arbitrarily structured laser light[36–39].

**Acknowledgements**
T.T., H.L., A.D.A., H.B. and N.M. acknowledge support from the Swedish Research Council (grant no. 2021-05784), Kempestiftelserna (grant no. JCK-3122) and the European Innovation Council (grant n. 101046920). S.W.J. acknowledge support from the Fonds de la Recherche Scientifique-FNRS.

[26] G. Milione, H.I. Sztul, D.A. Nolan, and R.R. Alfano, "Higher-Order Poincaré Sphere, Stokes Parameters, and the Angular Momentum of Light," Phys. Rev. Lett. **107**(5), 053601 (2011).

[27] G. Milione, S. Evans, D.A. Nolan, and R.R. Alfano, "Higher Order Pancharatnam-Berry Phase and the Angular Momentum of Light," Phys. Rev. Lett. **108**(19), 190401 (2012).

[28] J. Yao, X. Jiang, J. Zhang, A. Wang, and Q. Zhan, "Quantitative detection of high-order Poincaré sphere beams and their polarization evolution," Opt. Express **31**(2), 3017 (2023).

[29] A. Borot, and F. Quéré, "Spatio-spectral metrology at focus of ultrashort lasers: a phase-retrieval approach," Opt. Express **26**(20), 26444 (2018).

[30] C. Iaconis, and I.A. Walmsley, "Self-referencing spectral interferometry for measuring ultrashort optical pulses," IEEE J. Quantum Electron. **35**(4), 501–509 (1999).

[31] J. Pan, Y. Chen, Z. Huang, C. Zhang, T. Chen, D. Liu, D. Wang, M. Pang, and Y. Leng, "Self-Referencing 3D Characterization of Ultrafast Optical-Vortex Beams Using Tilted Interference TERMITES Technique," Laser & Photonics Reviews **17**(4), 2200697 (2023).

[32] M.A. Porras, "Upper Bound to the Orbital Angular Momentum Carried by an Ultrashort Pulse," Phys. Rev. Lett. **122**(12), 123904 (2019).

[33] M.A. Porras, and S.W. Jolly, "Control of vortex orientation of ultrashort optical pulses using spatial chirp," (2023).

[34] A. Grupp, A. Budweg, M.P. Fischer, J. Allerbeck, G. Soavi, A. Leitenstorfer, and D. Brida, "Broadly tunable ultrafast pump-probe system operating at multi-kHz repetition rate," Journal of Optics **20**(1), 014005 (2017).

[35] T. Tapani, H. Lin, N. Henriksson, and N. Maccaferri, in *Nonlinear Optics and Applications XIII*, edited by A.V. Zayats, M. Bertolotti, and A.M. Zheltikov (SPIE, Prague, Czech Republic, 2023), p. 7.

[36] G. Molina-Terriza, J.P. Torres, and L. Torner, "Management of the Angular Momentum of Light: Preparation of Photons in Multidimensional Vector States of Angular Momentum," Phys. Rev. Lett. **88**(1), 013601 (2001).

[37] A. Forbes, "Structured Light from Lasers," Laser & Photonics Reviews **13**(11), 1900140 (2019).

[38] M. Piccardo, M. De Oliveira, A. Toma, V. Aglieri, A. Forbes, and A. Ambrosio, "Vortex laser arrays with topological charge control and self-healing of defects," Nat. Photon. **16**(5), 359–365 (2022).

[39] J.R. Pierce, J.P. Palastro, F. Li, B. Malaca, D. Ramsey, J. Vieira, K. Weichman, and W.B. Mori, "Arbitrarily structured laser pulses," Phys. Rev. Research **5**(1), 013085 (2023).